# Deep Learning to Address Candidate Generation and Cold Start Challenges in Recommender Systems: A Research Survey

Kiran R, Pradeep Kumar, Bharat Bhasker
(efpm04013@iiml.ac.in, pradeepkumar@iiml.ac.in, bhasker@iimraipur.ac.in)
IT and Systems Department
Indian Institute of Management Lucknow
Lucknow 226013, India

Corresponding Author: Kiran Rama
Address: No 295, 3rd Block, 6th Cross, HMT Layout, Vidyaranyapura PO, Bangalore - 560097
Telephone : +91 9900601013
Email: efpm04013@iiml.ac.in

Author: Prof Pradeep Kumar
Indian Institute of Management Lucknow , Off Sitapur Road , Lucknow 226013, India

Author: Prof Bharat Bhasker
Indian Institute of Management Raipur, GEC Campus, Sejbahar, Raipur, Chhattisgarh 492015, India

# ABSTRACT

Among the machine learning applications to business, recommender systems would take one of the top places when it comes to success and adoption. They help the user in accelerating the process of search while helping businesses maximize sales. Post phenomenal success in computer vision and speech recognition, deep learning methods are beginning to get applied to recommender systems. Current survey papers on deep learning in recommender systems provide a historical overview and taxonomy of recommender systems based on type. Our paper addresses the gaps of providing a taxonomy of deep learning approaches to address recommender systems problems in the areas of cold start and candidate generation in recommender systems. We outline different challenges in recommender systems into those related to the recommendations themselves (include relevance, speed, accuracy and scalability), those related to the nature of the data (cold start problem, imbalance and sparsity) and candidate generation. We then provide a taxonomy of deep learning techniques to address these challenges. Deep learning techniques including convolutional neural networks, recurrent neural networks, autoencoders, stacked multi-layer perceptron, stacked Restricted Boltzmann Machines and Generative Adversarial Networks are categorized as supervised and unsupervised, as well as based on their applicability to different types of data. These techniques are mapped to the different challenges in recommender systems providing an overview of how deep learning techniques can be used to address them. We contribute a taxonomy of deep learning techniques to address the cold start and candidate generation problems in recommender systems. Cold Start is addressed through additional features (for audio, images, text) and by learning hidden user and item representations. Candidate generation has been addressed by separate networks, RNNs, autoencoders and hybrid methods. We also summarize the advantages and limitations of these techniques while outlining areas for future research.

# 1 INTRODUCTION

The business world today is characterized by information overload and increased competition for customer's attention and wallet share. In this context, Recommender Systems provide several benefits to both consumers and business firms. They help consumers with reduced search costs, increased product fit and management of choice overload. They help business firms with higher sales volume, higher web usage, higher customer retention and higher margins (Lee, Huntsman, Fl, Huntsman, & Fl, 2014). Several successes of recommender systems have been highlighted like 80% of movies watched on Netflix came from recommendation (Gomez-Uribe & Hunt, 2015) and 60% of video clicks came from home page recommendations on YouTube (Davidson et al., 2010). We define Recommender systems based on several definitions in literature as

*"Systems that seek to predict the future preference of a set of items for a user either as a numeric rating for the item or as a list of recommendations or as a binary score indicating preference for the item"* (Handbook, RICCI, & ROKACH, 2011), (Portugal, Alencar, & Cowan, 2017) (Cheng et al., 2016) (S. Zhang & Yao, 2017)

Advances in graphical processing unit hardware and decreasing costs have made GPUs accessible. With their multi-core nature, they enable highly parallel matrix computations that are at the core of Deep Learning. GPUs have been found to speed up the learning of a deep learning network by a factor of more than 50 (Schmidhuber, 2015). Consequently, Deep Learning has seen tremendous success in computer vision and speech recognition. Deep Learning methods employ neural networks with multiple hidden layers and are shown to be infinitely flexible functions capable of solving any problem provided there is sufficient data and computing power. They fit these several thousands of parameters using gradient descent. There have been tremendous improvements in backpropagation and gradient descent since the seminal paper on gradient descent in 1986 (Rumelhart, Hinton, & William, 1985). There are several methods for optimization used like RMSProp (Dauphin, de Vries, & Bengio, 2015), AdaDelta, AdaMax, Nadam, AMSGrad, Adam (Kingma & Ba, 2015) etc. The advancement of several frameworks like PyTorch from Facebook (Ketkar, 2017), MXNet from Apache (Chen et al., 2015) and TensorFlow (Google, 2018) from Google have increased the adoption of deep learning.

The motivation to write this survey paper is to not to communicate the list of results but to provide an understanding into deep learning applications in recommender systems, providing a taxonomy of how deep learning techniques have been used to address the candidate generation and cold start challenges in recommender systems.

## 1.1 RESEARCH GAP AND OBJECTIVE

There exist several surveys in traditional recommender systems but to the best of our knowledge, there are very few surveys on the application of deep learning to recommender systems. (S. Zhang & Yao, 2017) classified the techniques into methods that solely relied on deep learning and those that integrate deep learning with traditional recommender systems, further breaking them as loosely coupled and tightly coupled. The paper provides a historical course of development of techniques in the field. Another classification of deep learning for recommender systems is bucketing them into content, collaborative and hybrid buckets (Singhal, Sinha, & Pant, 2017). (Betru & Onana, 2017) is another survey paper on deep learning in recommender systems. None of these survey papers provide a taxonomy of deep learning methods based on the challenges of recommender systems that they address.

The research objective of this paper is to address the literature gaps mapping different deep learning techniques against the recommendation system challenges they address and provide a mapping of the techniques into applicable scenarios. The scope of this paper will be limited to the application of deep learning techniques to address the cold start and candidate generation problems in recommender systems.

## 1.2 RESEARCH METHODOLOGY

We survey different research papers on the application of deep learning to recommender systems. We extract articles from Google Scholar Search for the period 2009-present using combinations of the following keywords: deep learning, recommender systems, rec sys, recurrent neural networks, convolutional neural networks, stacking Restricted Boltzmann Machines, Deep Neural Networks, Autoencoders, Candidate Generation and Cold start. The resulting papers are manually curated based off the paper title and abstracts to find the articles. The deep learning techniques in the papers are then mapped to recommender system challenges building out a taxonomy. We avoid taking the historical course of development approach or the taxonomy based on content-based and collaborative filtering methods as these have already been addressed by several other survey papers. We focus on the techniques used to address specific challenges, their advantages and limitations and areas for further research.

## 1.3 ORGANIZATION OF THE PAPER

In the first section, we introduce recommender systems and deep learning topics. We also highlight the research objective and methodology. In the second section, we provide a background to the recommender system challenges and deep learning by providing a taxonomy. In the next section, we contribute to the research on the topic by coming up with a taxonomy of deep learning methods to handle the problems of cold start and candidate generation in recommender systems. In the last two sections, we provide a summary of the limitations of the research in the area and conclude by summarizing our contributions and areas for future research.

# 2   BACKGROUND

In this section, we summarize the challenges associated with recommender systems. We then provide a taxonomy of the deep learning methods, both based on the technique and based on the area of applicability.

## 2.1   RECOMMENDER SYSTEM CHALLENGES

The challenges in recommender systems can be classified into those related to the recommendations themselves, nature of the data and candidate generation. This is summarized in Figure 2.1. Challenges related to the recommendations themselves include speed, scalability, accuracy and relevance. Speed and Scalability are satisficing criteria while accuracy and relevance are optimizing criteria. Recommender systems are expected to provide the outputs in a reasonable amount of time while scaling to large amounts of data. Accuracy is measured by different metrics based on the nature of the task like RMSE, RMSLE (for regression problems), F-score, AUC, accuracy (for classification problems), NDCG and MAP (for ranking problems). The key goal here in relevance to avoid noise and provide fresh recommendations. Challenges related to the nature of the data include cold start problem, imbalance and sparsity. When there is no history about the user to whom the recommendation is to be made or the item that is to be recommended, the recommender systems run into the 'cold start' problem. Funnily, the origin of the cold start problem was from cars in cold regions that had problems starting up and running till it reaches its optimum temperature. For recommender systems, this means that there are not enough historical interactions for item or user.

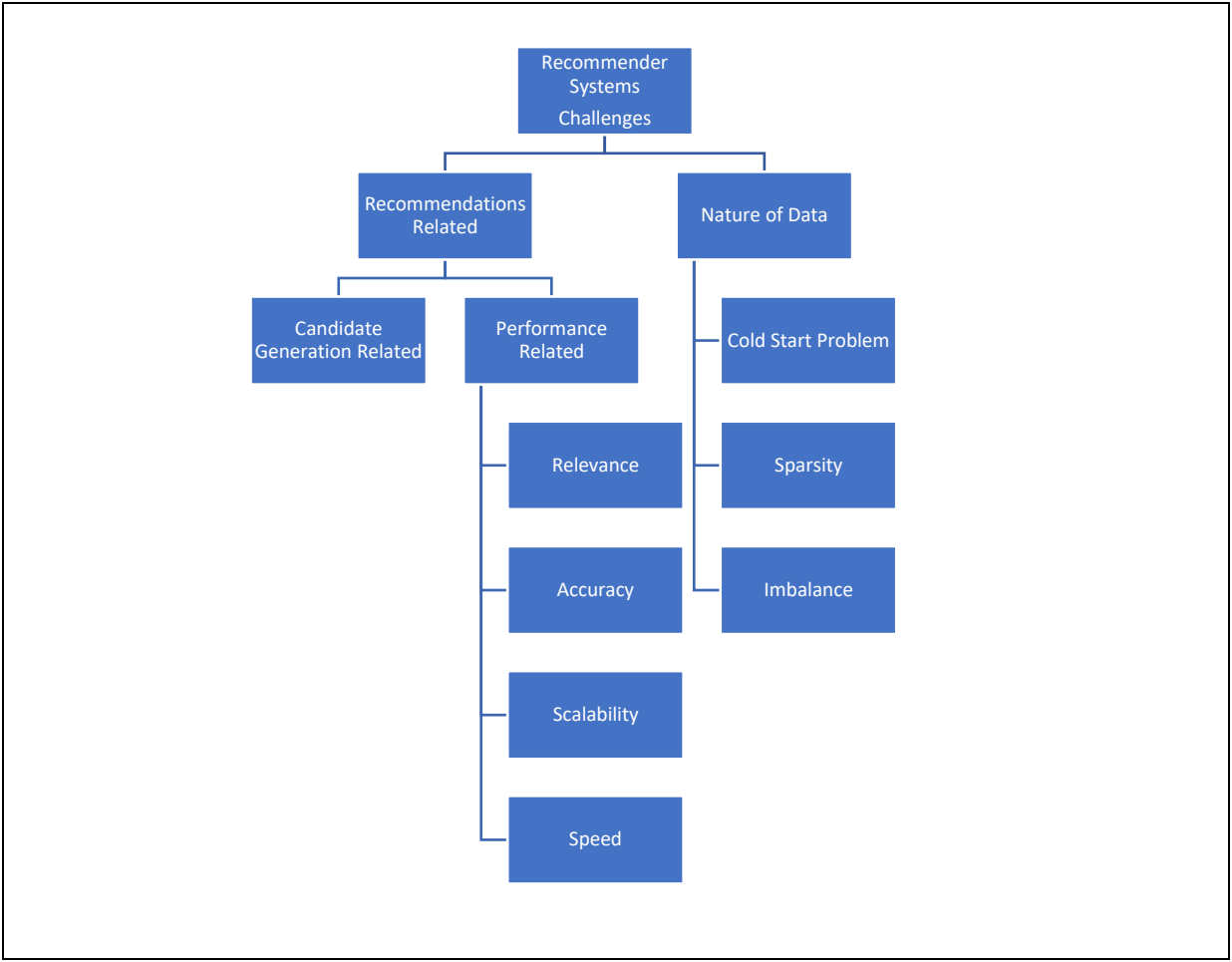

Fig 2.1: Challenges in Recommender Systems

Larger share of the ratings matrix is empty with no ratings for a user-item pair and this leads to the sparsity problem. Very few item-user combinations in the data matrix are minority class examples leading to the problem of imbalance. The last bucket of problem relates to candidate generation. Creating an entry for all user-item combinations is an onerous task and it becomes necessary to generate pairs of users and items for training. For example: With 1 million videos and 100 million users, YouTube has a candidate generation problem as it cannot create an entry for every user-video pair. (Covington, Adams, & Sargin, 2016).

## 2.2 TAXONOMY OF DEEP LEARNING METHODS

In figure 2.2, we categorize the deep learning techniques based on their complexity and type.

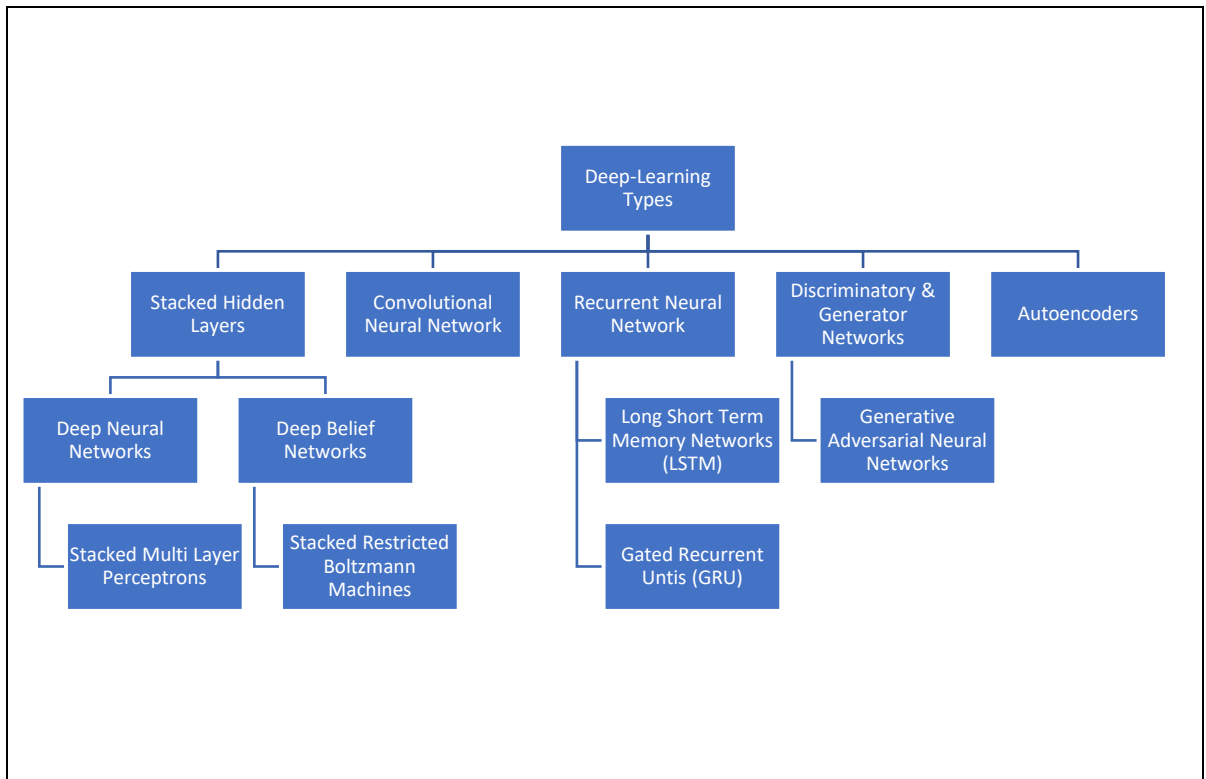

Figure 2.2: Taxonomy of Deep Learning Methods by Method and Architecture Type

Multi-layer perceptrons (MLP) are feedforward neural networks with multiple hidden layers between the input layer and the output layer. Restricted Boltzmann Machines (RBM) are two layered neural networks with a visible layer and a hidden layer. Both MLPs and RBMs are traditional neural networks. By stacking many RBNs together, we get a Deep Belief Network and by stacking several hidden layers in a MLP together, we get a Deep Neural Network (DNN). CNNs have several convolutional layers that apply filters followed by pooling layers with non-linear transformations and have found huge success in image processing. RNNs are suitable for modelling sequential recommendations and they are a unique type of neural network that contain loops that enable the networks to have memory storage. RNNs are of two types namely Long Short Term Memory (LSTM) and Gated Recurrent Units (GRU). Generative Adversarial

Networks (GANs) are generative neural networks that consist of a discriminator and a generator network with both trained simultaneously competing with one another. All the methods discussed so far are supervised deep learning methods. Autoencoders are unsupervised models that attempt to reconstruct the input data in the output layer and use the bottleneck layer as a salient representation of the input. Figure 2.3 shows the classification of the deep learning methods based on the popular types of data they are applied to. It is to be noted that these are not the "only" applications but rather the most popular applications

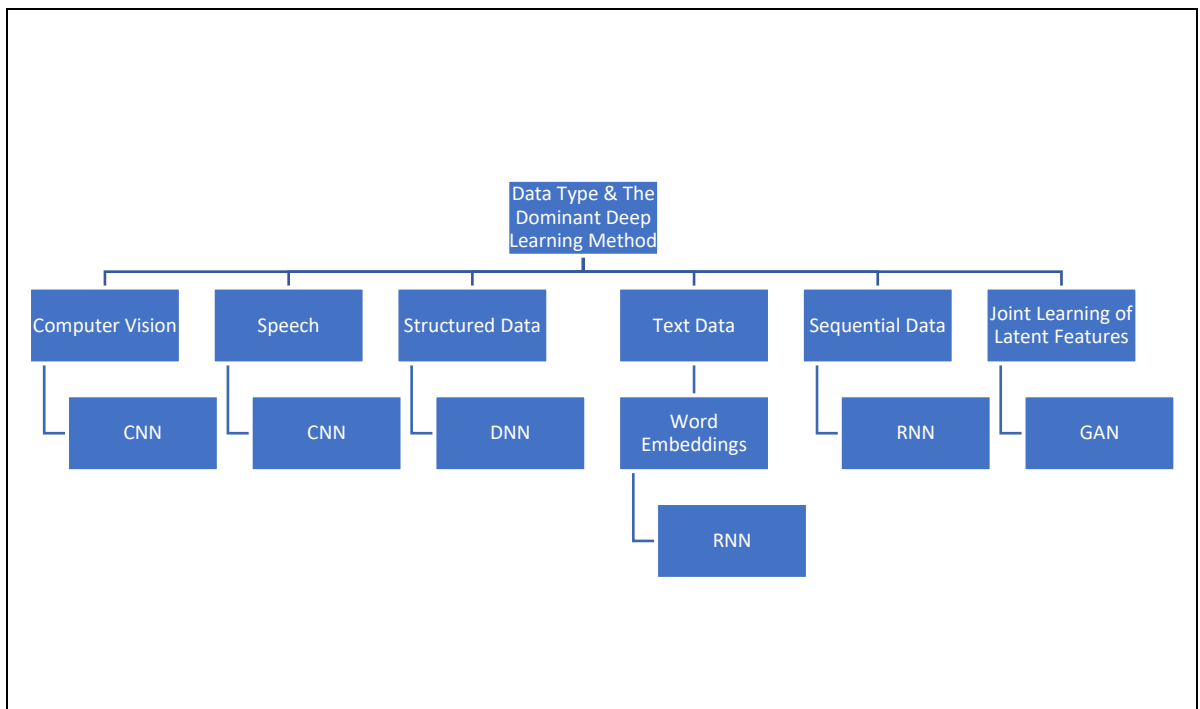

Figure 2.3: Taxonomy of Deep Learning Methods by Popular application type

# 3 PROPOSED TAXONOMY OF DEEP LEARNING TECHNIQUES TO ADDRESS COLD START AND IMBALANCE

In the section, we map the different deep learning techniques and variants to the recommender system challenges. We then summarize the advantages and disadvantages of each of these approaches.

## 3.1 COLD START

Cold Start Problem is one of "Making recommendations where there are no prior interactions available for an user or an item" (Lam, Vu, & Le, 2008). (Bernardi, Kamps, Kiseleva, & Mueller, 2015; Lika, Kolomvatsos, & Hadjiefthymiades, 2014) highlight the cold start problem and classify it into user cold start and item cold start cases, referring to cases of insufficient examples of items and users respectively.

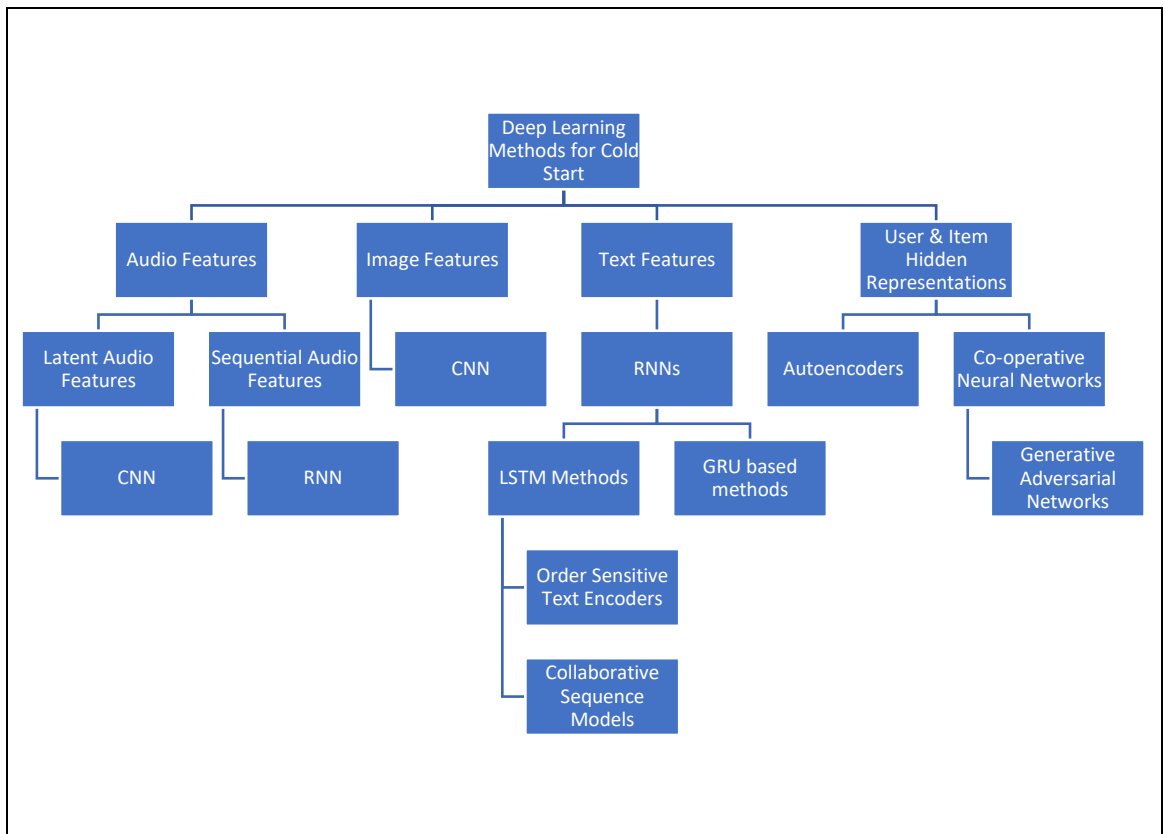

Fig 3.1: Taxonomy of Deep Learning Methods to address Cold Start Problem

The usual approach to solving the cold start problem is to include additional user features by combining content based and collaborative filtering based methods (Chiliguano & Fazekas, 2016).

Modern Song Recommender Systems fail to recommend tracks that are less widely known and suffer from the cold start problem. Deep convolutional networks have been used to generate latent factors for songs from their audio when no usage is available (Oord, Dieleman, & Schrauwen, 2013) and using audio content of the song to add some item based features alleviates the cold start problem. CNNs have been applied to extract features from images (Oord et al., 2013). The sequential property of audio has been used to build a LSTM-based recurrent neural network architecture in combination with audio based features from a convolutional networks have been attempted. (Balakrishnan, 2014). Content-Based Deep Learning methods include extracting latent factors for songs from their audio (Liang, 2014), enhancing content based quote recommendations(Tan, Wan, & Xiao, 2016), converting item text into latent features. GRU based recurrent neural networks have also been used to convert item text into latent features to improve the collaborative filtering results for the cold start problem (Bansal, Belanger, & McCallum, 2016).

For textual data, order sensitive encoders in combination with latent factor models have been used to address the cold start problem by employing RNNs to encode text sequences into latent factors (Bansal et al., 2016). Extracted item representations are combined with user embeddings to get predicted ratings for each user-item pair. Compared to traditional methods that ignore word order, leveraging the word order in recurrent neural networks helps soften the cold start problem. Collaborative Sequence models based on recurrent neural networks have also shown good results. (Ko, Maystre, Grossglauser, Durrant, & Kim, 2016). (Devooght & Bersini, 2016) show that collaborative filtering can be viewed as a sequence prediction problem and application of LSTMs is very competitive in addressing the cold start problem.

Latent representations from the data are learnt in an unsupervised manner and the implicit relationships between items and users are learnt from both the content and the rating (Xiaopeng Li & She, 2017). The autoencoder is forced to learn a latent distribution for content in latent space instead of observation space through an inference network and can easily be extended to non-text type multimedia as well. Instead of learning the user latent factors and item latent factors separately, Co-operative Neural Networks (CoNN) learn hidden latent features for users and items jointly using two coupled neural networks. These are examples of Generative Adversarial Networks (GANs) which consist of a discriminator and a generator networks and both are trained simultaneously competing in a minimax game framework

## 3.2  CANDIDATE GENERATION

If there are 1 million users and 0.3 million items, a data matrix at a user-item level would have $0.3 * 10^{12}$ combinations as an input into a data mining algorithm to solve a classification or prediction task. Such a huge data matrix lends itself unviable from a computational and imbalance perspective. Hence candidate generation is an important problem in recommender systems. A summary of the methods in literature to solve the problem of candidate generation in deep learning is shown in Figure 3.2

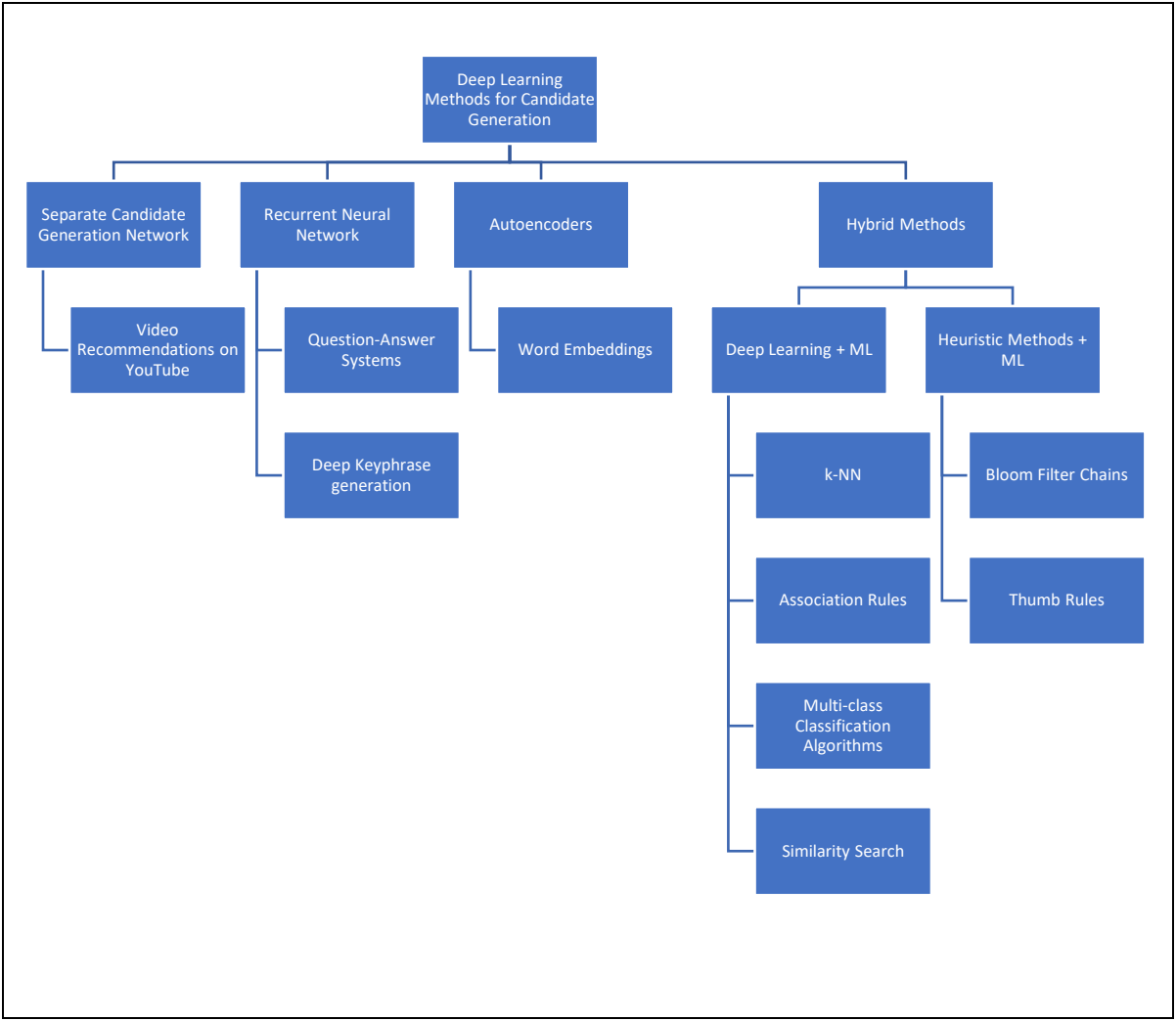

Fig 3.2: Taxonomy of Deep Learning Methods to address Candidate Generation Problem

YouTube used a candidate generation network (Covington et al., 2016) with inputs as events from user history. The problem of returning hundreds of videos from millions was treated as an extreme multi-class problem and is solved using a deep neural network that learnt the user embeddings as a function of the user's history and context. RNNs have been used for candidate generation for sequential data in Question-Answer Systems and deep keyphrase generation. In (Shen, Chen, & Huang, 2016), the authors turn to paraphrases as a means of capturing the way natural language expresses the information needed and present a general framework using a deep neural model which learns felicitous paraphrases for various tasks. In (Y. Zhang, Fang, &

Weidong, 2018), first a list of phrase candidates with heuristic methods are returned followed by scoring each candidate phrase for its likelihood of being a keyphrase in the given document. (Xiaonan Li, Li, & Yu, 2012) combined word embeddings with L2 regularized large scale logistic regression to search sentences similar to a query from Wikipedia articles and directly use the human-annotated entities in the similar sentences as candidate entities for the query.

Trivial business rule-based techniques and simple machine learning methods have been used combined with deep learning extensively in literature to create hybrid systems to solve the problem of candidate generation. In Google Apps, there are over a million apps in the database. Every app cannot be scored for every query. Returning a short list of items that best match the query using various signals is achieved using a combination of machine learned methods and human defined rules (Cheng et al., 2016). Bloom Filter Chains have also been used for candidate generation for real-time tweet search involving sorting of tweets using a simple scoring function and then rescoring them with a separate component (Asadi & Lin, 2013). IBM Watson, a commercial tool calls out "Hypothesis Generation" as a key component and the same uses candidate generation to retrieve answers to a particular question (Chu-Carroll et al., 2012). Association Rules, a classic machine learning technique has also been used for candidate generation as Generalized Sequential Patterns (Srikant & Agrawal, 1996). Traditional methods have also been combined with deep learning techniques to create hybrid systems.

# 4   OPPORTUNITIES

On candidate generation, while there have been novel implementations of deep learning, they have been very domain focused like candidate generation network using Deep Neural Network for YouTube, RNNs for Question-Answer Systems and Deep Keyphrase generation and autoencoders for word embeddings. These methods have been shown to solve the candidate generation problem in a specific domain, but they do not generalize to all scenarios. Also, the candidate generation methods apart from the YouTube example have not combined deep learning with traditional machine learning techniques like k-NN, association rules, bloom filters and classification algorithms. These present an opportunity for future research.

On cold start, hand-crafting features has been reduced greatly in the cases of images, audio and text where CNNs and RNNs have shown tremendous results. Still, for traditional structured recommendations data, hand-crafting features is required in many cases, which means that features need to be maintained and domain experts need to be deployed. Employing User and item features effectively into deep network in an automated way into deep learning methods to solve the cold start problem is very important. Common to both cold start and candidate generation, there are very few examples of using the different nature of deep learning methods in a way complementary to one another. Some examples include combining RNNs and autoencoders. Deep Neural Networks on text ignore the word order and semantic meanings but Recurrent Neural Networks can take the word order into account. Autoencoders can leverage the semantic order. Combining these methods can potentially give better results. The actual incremental improvement that can be expected from deep learning methods on candidate generation remains an important research area. An immediate improvement in alleviating cold start becomes possible where there have been several implementations taking one of the techniques into account. In one of the papers, latent factors learned from Weighted Matrix Factorization was not accurate enough to train a CNN, but could have easily adapted to a different technique.

## 4.1 RESEARCH OPPORTUNITIES
Some of the research areas that we have identified include:

- Using Autoencoders to effectively automatically engineer features to solve the cold start problem. The autoencoder approach to feature engineering that converts a user vector to a dense vector could bring out a significant improvement in the performance on the experimental datasets, but there is not much literature on it. Autoencoders seem to have been used heavily for text recommendations but not on structured datasets
- Factorization machines can avoid tedious feature engineering for cold start problem and could be promising to integrate them with deep learning techniques
- Exploring CNNs for structured data. Techniques from image processing to features like filters, pooling, residual networks, one-pass object detections could find parallels in structured data modeling. This is in addition to leveraging CNNs for extracting features from images data
- Recommendations are a form of sequential data. There is not enough literature on recurrent neural networks and their types LSTM and GRU to automate some feature engineering
- All algorithms have tried to optimize for overall accuracy metrics like RMSE, RMSLE, AUC, F-score, NDCG, MAP etc., which are global measures for the whole dataset. There could be potential in coming up with a custom evaluation metric that gives higher weight to cold start cases

- Coming up with hybrid methods that combine Deep learning with Association rules or k-NN for a generalized framework for candidate generation is a promising area

# 5  CONCLUSION

Current survey papers on deep learning in recommender systems provide a historical overview and taxonomy of recommender systems based on type. We have grouped the recommender system challenges into those related to recommendations, those related to candidate generation and those related to the nature of the data. We have focused on candidate generation and cold start challenges in this paper. Our paper addresses the gaps of providing a taxonomy of deep learning approaches to address recommender systems problems in the areas of cold start and candidate generation. Cold Start problem is due to insufficient information about users or items. Deep Learning techniques have tried to address this by generating features for the audio (CNN, RNN), for the images (CNN), for the text (LSTM methods and GRU methods) and for hidden user and item representations (Autoencoders and GANs). The candidate generation problem has been addressed by creating separate candidate generation networks (DNN) and recurrent neural networks. We have summarized the limitations of each of the methods and the important research opportunities

Our key contribution in this survey paper is the taxonomy of deep learning methods to address cold start and candidate generation with a summary of the limitations and opportunities for research